\documentclass[aps,prl,twocolumn,superscriptaddress,showpacs]{revtex4}

\usepackage{graphicx}
\usepackage{amsmath}

\begin{document}

\title{Momentum-Resolved Tunneling into Fractional Quantum Hall
       Edges}

\author{U. Z\"ulicke}
\affiliation{Institut f\"ur Theoretische Festk\"orperphysik,
Universit\"at Karlsruhe, D-76128 Karlsruhe, Germany}

\author{E. Shimshoni}
\affiliation{Department of Mathematics and Physics, University of
Haifa at Oranim, Tivon 36006, Israel}

\author{M. Governale}
\affiliation{Institut f\"ur Theoretische Festk\"orperphysik,
Universit\"at Karlsruhe, D-76128 Karlsruhe, Germany}

\date{\today}

\begin{abstract}

Tunneling from a two--dimensional contact into quantum--Hall edges is
considered theoretically for a case where the barrier is extended,
uniform, and parallel to the edge. In contrast to previously realized
tunneling geometries, details of the microscopic edge structure are
exhibited directly in the voltage and magnetic--field dependence of
the differential tunneling conductance. In particular, it is possible
to measure the dispersion of the edge--magnetoplasmon mode, and the
existence of additional, sometimes counterpropagating,
edge--excitation branches could be detected.

\end{abstract}


\maketitle

The quantum Hall (QH) effect\cite{qhe-sg} arises due to
incompressibilities developing in two--dimensional electron
systems (2DES) at special values of the electronic sheet density
$n_0$ and perpendicular magnetic field $B$ for which
the {\em filling factor\/} $\nu = 2\pi\hbar c\, n_0/|e B|$ is 
equal to an integer or certain fractions. The
microscopic origin of incompressibilities at fractional $\nu$ is
electron--electron interaction. Laughlin's trial--wave--function
approach~\cite{rbl:prl:83} successfully explains the QH effect at
$\nu=\nu_{1,p}\equiv 1/(p+1)$ where $p$ is a positive even
integer. Our current microscopic understanding of why
incompressibilities develop at many other fractional values of the
filling factor, e.g., $\nu_{m,p}\equiv m/(m p + 1)$ with nonzero
integer $m\ne\pm 1$, is based on hierarchical
theories~\cite{fdmh:prl:83,bih:prl:84,jain:prl:89}.

The underlying microscopic mechanism responsible for creating
charge gaps at fractional $\nu$ implies peculiar properties of
low--energy excitation in a finite quantum--Hall sample which are
localized at the boundary~\cite{ahmintro}. For $\nu=\nu_{m,p}$,
$m$ branches of such edge 
excitations~\cite{ahm:prl:90,wen:prb:90,wen:int:92,wen:adv:95}
are predicted to exist which are realizations of strongly
correlated chiral one--dimensional electron systems called {\em
chiral Luttinger liquids\/} ($\chi$LL). Extensive experimental
efforts were undertaken recently to observe $\chi$LL behavior
because this would yield an independent confirmation of our basic
understanding of the fractional QH effect. In all of these
studies\cite{webb:ssc:96,amc:prl:96,amc:prl:98,amc:prl:01,matt:prl:01,hilke:prl:01},
current--voltage characteristics yielded a direct measure of the
energy dependence of the {\em tunneling density of states\/} for the
QH edge. This quantity generally contains information on global
dynamic properties as, e.g., excitation gaps and the orthogonality
catastrophe, but lacks any momentum resolution. Power--law behavior
consistent with predictions from $\chi$LL theory was
found\cite{webb:ssc:96,amc:prl:96,matt:prl:01} for the edge of QH
systems at the Laughlin series of filling factors, i.e., for $\nu=
\nu_{1,p}$. However, at hierarchical filling factors, i.e., when
$\nu=\nu_{m,p}$ with $|m|>1$, predictions of $\chi$LL theory are,
at present, not supported by
experiment~\cite{amc:prl:98,amc:prl:01}. This discrepancy inspired
theoretical works, too numerous to cite here, from which, however,
no generally accepted resolution emerged. Current
experiments\cite{hilke:prl:01} suggest that details of the edge
potential may play a crucial r\^ole. New experiments are needed to
test the present microscopic picture of fractional--QH edge
excitations.

\begin{figure}
\includegraphics[width=3.2in]{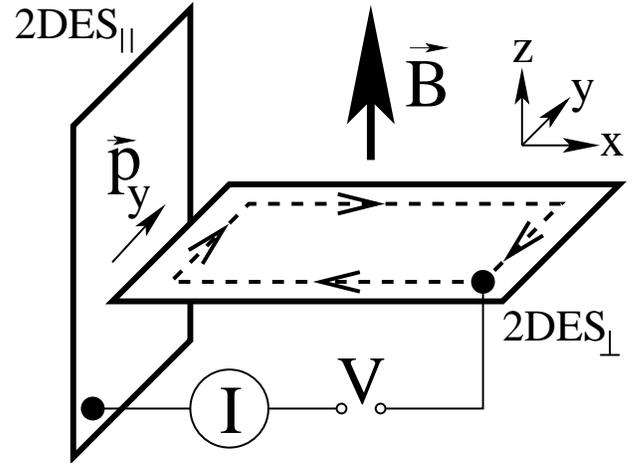}
\caption{Schematic picture of tunneling geometry. Two mutually
perpendicular two-dimensional electron systems are realized, e.g.,
in a semiconductor heterostructure. An external magnetic field is
applied such that it is perpendicular to one of them
(2DES$_\perp$) but in-plane for the other one (2DES$_\parallel$).
When 2DES$_\perp$ is in the quantum-Hall regime, chiral edge
channels form along its boundary (indicated by broken lines with
arrows). Where they run parallel to 2DES$_\parallel$, electrons
tunnel between edge states in 2DES$_\perp$ and plane-wave states
in 2DES$_\parallel$ with the {\em same\/} quantum number $p_y$ of
momentum component parallel to the barrier. Experimentally, the
differential tunneling conductance $dI/dV$ is measured.
\label{setup}} 
\end{figure}                  

Here we consider a tunneling geometry that is particularly well-suited
for that purpose, see Fig.~\ref{setup}, and which has been realized
recently for studying the integer QH effect in cleaved-edge overgrown
semiconductor hetero\-structures\cite{matt:physe:02}. In contrast to
previous experiments, it provides a {\em momentum--resolved} spectral
probe of QH edge excitations\footnote{Tunneling from a {\em
three--dimensional\/} contact into a QH edge, measured in
Refs.~\onlinecite{amc:prl:96,amc:prl:98,amc:prl:01,matt:prl:01,hilke:prl:01},
cannot resolve momentum even with perfect translational invariance
parallel to the edge. The latter is destroyed anyway, in real samples,
by dopant--induced disorder in the bulk contact. See also a related
tunneling spectroscopy of parallel QH edges by W. Kang {\it et al.},
Nature (London) {\bf 403}, 59 (2000).}. With both the component of
canonical momentum parallel to the barrier and energy being conserved
in a single tunneling event, strong resonances appear in the
differential tunneling conductance $dI/dV$ as a function of the
transport voltage and applied magnetic field. Similar resonant
behavior for tunneling via extended uniform barriers has been used
recently\cite{jpe:apl:91,cav:prl:96,eaves:sci:00,ophir:sci:02} to
study the electronic properties of low--dimensional electron
systems. It has also been suggested as a tool to observe
spin--charge separation in Luttinger liquids\cite{hekk:prl:99} and
the interaction--induced broadening of electronic spectral
functions at single-branch QH edges\cite{uz:prb-rc:96}. Here we
find that the number of resonant features in $dI/dV$ corresponds
directly to the number of chiral edge excitations present.
Edge--magnetoplasmon dispersion curves can be measured and power
laws related to $\chi$LL behavior be observed. Momentum--resolved
tunneling spectroscopy in the presently considered geometry thus
constitutes a powerful probe to characterize the QH edge
microscopically.

\begin{figure}
\includegraphics[width=3.3in]{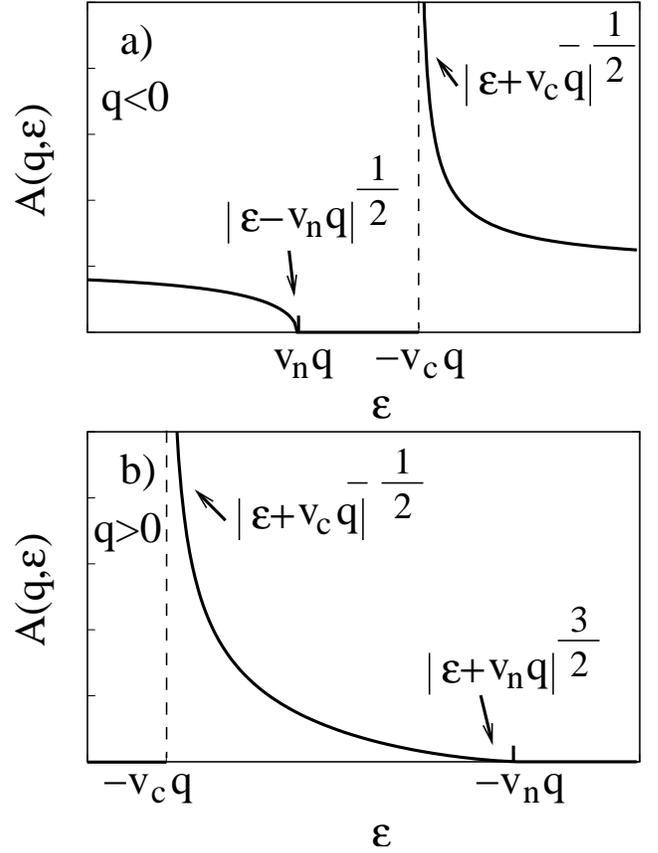}
\caption{Spectral functions for two--branch hierarchical
fractional--QH edges at bulk filling factor $2/3$ [panel a)] and
$2/5$ [panel b)], where the charged (edge--magnetoplasmon) mode is
assumed to be left--moving. a)~We show $A_{2/3}^{(0)}(q,\varepsilon
)\equiv A_{2/3}^{(1)}(q,\varepsilon)$ for a fixed value of $q$.
Note the similarity with the spectral function of a spinless
Luttinger liquid\cite{med:prb:92,voit:prb:93}. The only difference
is that, in our case, velocities of right--moving and left--moving
plasmon modes are not equal. b)~$A_{2/5}^{(0)}(q,\varepsilon)\equiv
A_{2/5}^{(-1)}(q,\varepsilon)$ at fixed $q$. It is reminiscent of
the spectral function for a spinful $\chi$LL exhibiting
spin--charge separation\cite{fab:prl:93,voit:prb:93} but differs
due to the absence of any algebraic divergence at $-v_{\text{n}}q$.
\label{spectr}} 
\end{figure}                  

To compute the tunneling conductances, we apply
the general expression for the current obtained to lowest order in
a perturbative treatment of tunneling\cite{mahan}:
\begin{eqnarray}\label{geniv}
I(V)&=&\frac{e}{\hbar^2}\sum_{\vec k_\parallel, n, X} |t_{\vec
k_\parallel,n,X}|^2 \int\frac{d\varepsilon}{2\pi}\left\{
n_{\text{F}}(\varepsilon)-n_{\text{F}}(\varepsilon+eV)\right\}
\nonumber \\ && \hspace{2cm}\times A_\parallel (\vec k_\parallel,
\varepsilon) \, A_\perp(n, X, \varepsilon+eV) \, .
\end{eqnarray}
Here $A_\parallel$ and $A_\perp$ denote single-electron spectral
functions for 2DES$_\parallel$ and 2DES$_\perp$, respectively. (See
Fig.~\ref{setup}).
We use a representation where electron states in the first are
labeled by a two--dimensional wave vector\footnote{Here we
neglect magnetic--field--induced subband mixing in
2DES$_\parallel$. While this can be straightforwardly included, it
will typically result in small quantitative changes only.}
$k_\parallel=(k_y, k_z)$, while the quantum numbers of electrons in
2DES$_\perp$ are the Landau--level index $n$ and guiding--center
coordinate $X$ in $x$ direction. We assume that 2DES$_\parallel$ is
located at $x=0$. The simplest form of the tunneling matrix element
$t_{\vec k_\parallel,n,X}$ reflecting translational invariance in
$y$ direction yields
\begin{equation}
t_{\vec k_\parallel,n,X}=t_n(X)\,\,\delta (k_y-k)\quad ,
\end{equation}
where $k\equiv X/\ell^2$ with the magnetic length $\ell=\sqrt{\hbar
c/|e B|}$. The dependence of $t_n(X)$ on $X$ results form the fact
that an electron from 2DES$_\perp$ occupying the state with quantum
number $X$ is spatially localized on the scale of $\ell$ around $x=
X$. The overlap of its tail in the barrier with that of states from
2DES$_\parallel$ will drop precipitously as $X/\ell$ gets large.
Finally, $n_{\text{F}}(\varepsilon)=[\exp(\varepsilon/k_{\text{B}}
T)+1]^{-1}$ is the Fermi function. In the following, we use the
expression $A_\parallel(\vec k_\parallel,\varepsilon)=2\pi\delta(
\varepsilon-E_{\vec k_\parallel})$ which is valid for a clean
system of noninteracting electrons\footnote{Broadening due to
scattering from disorder or interactions can be straightforwardly
included and does not change the main results of our study.}. Here
$E_{\vec k_\parallel}$ denotes the electron dispersion in
2DES$_\parallel$.

The spectral function of electrons in 2DES$_\perp$ depends crucially
on the type of QH state in this layer. At integer $\nu$, when
single--particle properties dominate and disorder broadening is
neglected, it has the form
\begin{equation}\label{intspec}
A_\perp(n,X,\varepsilon)\equiv A_n(k,\varepsilon)=2\pi\delta(
\varepsilon-E_{nk})\quad ,
\end{equation}
where $E_{nk}$ is the Landau--level dispersion. Strong correlations
present at fractional $\nu$ alter the spectral properties of edge
excitations. In the low--energy limit, it is possible to linearize
the lowest--Landau--level dispersion around the Fermi point
$k_{\text{F}}$. At the Laughlin series $\nu=1/(p+1)$ and for
short-range interactions present at the edge, the spectral function
was found\cite{wen:int:92,jujo:prl:96} to be
\begin{equation}\label{laughspec}
A_{\frac{1}{p+1}}(q,\varepsilon) = \frac{z}{p!}
\left(\frac{q}{2\pi/L_y}\right)^p \delta\left(\varepsilon-r
\hbar v_{\text{e}}q\right) .
\end{equation}
Here $q\equiv k-k_{\text{F}}$, $r=\pm$ distinguishes the two
chiralities of edge excitations, $L_y$ is the edge perimeter,
$v_{\text{e}}$ the edge--magnetoplasmon velocity, and $z$ an unknown
normalization constant. The power--law prefactor of the
$\delta$--function in Eq.~(\ref{laughspec}) is a manifestation of
$\chi$LL behavior.

The main focus of our work is the sharp QH edge at hierarchical
filling factors. Here we provide explicitly the momentum--resolved
spectral functions for $\nu=\nu_{\pm 2,p}$\footnote{Generalization
to $|n|>2$ is possible but does not add qualitatively new physical
insight.}. Microscopic theories\cite{ahm:prl:90,wen:int:92}
predict the existence of two Fermi points $k_{\text{Fo}}$ and
$k_{\text{Fi}}$ which correspond to outer and inner single-branch
chiral edges of QH fluids at Laughlin--series filling factors
$\nu_{\text{o}}^\pm=1/(p\pm 1)$ and $\nu_{\text{i}}^\pm=\pm 1/[(2p\pm
1)(p\pm 1)]$, respectively. The negative sign of $\nu_{\text{i}^-}$
indicates that the inner edge mode is counterpropagating. We have used
standard bosonization methods\cite{vondelft} applied to fractional--QH
edges\cite{wen:int:92} for the calculation of the spectral functions.
As these have not been obtained before, we briefly discuss their main
features here.

According to $\chi$LL theory, the existence of two Fermi points gives
rise to a discrete infinite set of possible electron tunneling
operators at the edge. This is because an arbitrary number $N$ of
fractional--QH quasiparticles with charge equal to $e
\nu_{\text{o}}^\pm$ can be transferred to the inner edge after an
electron has tunneled into the outer one\cite{wen:int:92}. Each of
these processes gives rise to a separate contribution to the
electronic spectral function at the edge which is of the general form
\begin{widetext}
\begin{eqnarray}
A^{(N)}_{\nu_{\pm 2,p}}(q,\varepsilon)&=&\frac{2\pi z}{\Gamma
(\eta^{(N)}_1)\Gamma(\eta^{(N)}_2)}\left(\frac{L_y/2\pi\hbar}{|v_1
\mp v_2|}\right)^{\eta^{(N)}_1+\eta^{(N)}_2-1}\left|\varepsilon-
r\hbar v_1 q \right|^{\eta^{(N)}_2-1}\left|\varepsilon\mp r\hbar
v_2 q \right|^{\eta^{(N)}_1-1}\nonumber \\ && \hspace{2cm}\times
\left\{\Theta\left(r\hbar v_1 q - \varepsilon\right)\Theta\left(\pm
\varepsilon-r\hbar v_2 q\right) + \Theta\left(\varepsilon - r\hbar
v_1 q\right)\Theta\left(r\hbar v_2 q\mp\varepsilon\right)\right\}.
\end{eqnarray}
\end{widetext}
Here $q\equiv k-k_{\text{F}}^{(N)}$, where $k_{\text{F}}^{(N)}=
k_{\text{Fo}}-N\nu_{\text{o}}^\pm(k_{\text{Fo}}-k_{\text{Fi}})$.
The velocities $v_1>v_2>0$ of normal--mode edge--density fluctuations
and the exponents $\eta_{1,2}^{(N)}$ depend strongly on microscopic
details of the edge, e.g., the self-consistent edge potential and
inter--edge interactions. We focus here on the experimentally
realistic case when inner and outer edges are strongly coupled and
the normal modes correspond to the familiar\cite{cllreview} charged
and neutral edge-density waves\footnote{Expressions for the general
case will be given elsewhere.}. In this limit, we
have\cite{cllreview,uz:prb:98} $v_1=v_{\text{c}}\sim {\mathcal O}
(\log[L_y/\ell])$, $v_2=v_{\text{n}}\sim {\mathcal O}(1)$ (where c
and n denote charged and neutral, respectively), and the exponents
assume universal values: $\eta_1^{(N)}=\eta_{\text{c}}\equiv p\pm1/
2$, $\eta_2^{(N)}=\eta_{\text{n}}^{(N)}\equiv(2 N\pm 1)^2/2$. Note
that exponents are generally larger than unity except for $N=0,\mp
1$ where $\eta_2^{(N)}=1/2$. In the latter case, an algebraic
singularity appears in the spectral function. This is illustrated
in Fig.~\ref{spectr}. Such divergences will be visible as strong
features in the differential tunneling conductance; see below.
Contributions to the spectral function for all other values of $N$
do not show such divergences and will give rise only to a featureless 
background in the conductance.

With spectral functions for 2DES$_\perp$ at hand, we are now able to
calculate tunneling transport. We focus first on the case when
2DES$_\perp$ is in the QH state at $\nu=1$. For realistic situations,
the differential tunneling conductance $dI/dV$ as a function of
voltage $V$ and magnetic field $B$ will exhibit two lines of strong
maxima whose positions in $V$--$B$ space are given by the equations
\begin{subequations}\label{edgedisp}
\begin{eqnarray}
\label{firstmax}
E_{0 k_V} &=& \varepsilon_{\text{F}\perp} \quad , \\
\label{secdmax}
E_{0 k_{\text{F}\parallel}} &=& \varepsilon_{\text{F}\perp} + e V 
\quad .
\end{eqnarray}
\end{subequations}
Here $k_V=\sqrt{2 m (\varepsilon_{\text{F}\parallel}- eV)/\hbar^2}$
and $k_{\text{F}\parallel}$, the Fermi wave vector in
2DES$_\parallel$, are the extremal wave vectors for which
momentum--resolved tunneling occurs. Fermi energies in 2DES$_{\perp,
\parallel}$ are denoted by $\varepsilon_{\text{F}\perp,\parallel}$.
Eqs.~(\ref{edgedisp}) can be used to extract the lowest--Landau--level
dispersion $E_{0 k}$ from maxima in the experimentally obtained $dI/dV
$, thus enabling microscopic characterization of real QH edges.

When 2DES$_\perp$ is in a QH state at a Laughlin--series filling
factor $\nu_{1,p}$, it supports a single branch of edge excitations
just like at $\nu=1$, and the calculation of the differential
tunneling conductance proceeds the same way. The major difference is,
however, the vanishing of spectral weight at the Fermi point of the
edge; compare Eqs.~(\ref{intspec}) and (\ref{laughspec}). This results
in the suppression of maxima described by Eq.~(\ref{firstmax}), while
those given by Eq.~(\ref{secdmax}) remain. The intensity of the latter
rises along the curve as a power law with exponent $p$.

\begin{figure}
\includegraphics[width=3.2in]{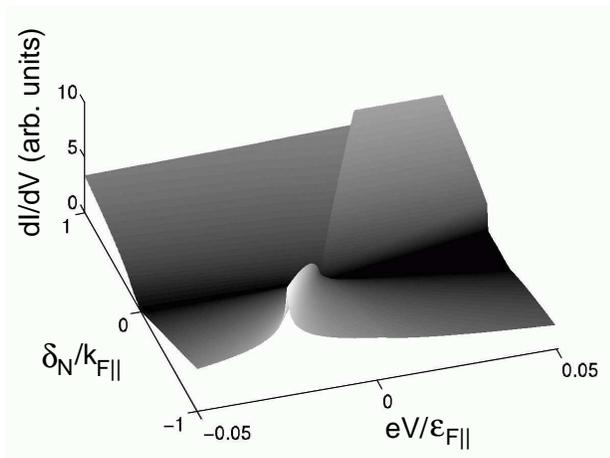}
\caption{Gray--scale plot of singular contributions to the
differential conductance for tunneling into the two--branch QH edge at
filling factor $2/3$. A qualitatively similar plot is obtained for
filling factor $2/5$. Note the strong maximum rising as a power law
for negative bias, which is continued as a step edge for positive
bias. Its position in the $eV$--$\delta_N$ plane follows a line whose
slope corresponds to the edge--magnetoplasmon velocity $v_{\text{c}}$.
To obtain the plot, we have linearized the spectrum in
2DES$_\parallel$ and absorbed the magnetic--field dependence into the
parameter $\delta_N=k_{\text{F}}^{(N)} - k_{\text{F}\parallel}$. As
there are two such singular contributions to $dI/dV$ with $N=0,1$
which have different $\delta_N$, a doubling of resonant features shown
in this plot would be observed experimentally.\label{twobranch}}
\end{figure}

Finally, we discuss the case of hierarchical filling factors $\nu_{\pm
2, p}$ which are expected to support two branches of edge excitations.
To be specific, we consider filling factors $2/3$ and $2/5$. In both
cases, there are many contributions to the spectral function and,
hence, the differential tunneling conductance. However, only two of
these exhibit algebraic singularities. It turns out that these
singularities give rise to either a strong maximum or a finite step in
the differential tunneling conductance, depending on the sign of
voltage. (See Fig.~\ref{twobranch}). The strong maximum results from a
logarithmic divergence that occurs when $eV=\hbar v_{\text{c}}(
k_{\text{F}}^{(N)}-k_{\text{F}\parallel})$. Both the maximum and the
step edge follow the dispersion of the charged edge--magnetoplasmon
mode and would therefore enable its experimental investigation. Most
importantly, however, the two spectral functions with singularities
exhibit them slightly shifted in guiding--center, i.e., $k$ direction
by an amount $\nu_{\text{o}}^\pm(k_{\text{Fo}}-k_{\text{Fi}})$. Hence,
two maxima and a double--step feature should appear in the
differential tunneling conductance whose distance in magnetic--field
direction will be a measure of the separation of inner and outer
edges. Observation of this doubling would yield an irrefutable
confirmation of the expected multiplicity of excitation branches at
hierarchical QH edges.

In conclusion, we have calculated the differential conductance for
momentum--resolved tunneling from a 2DES into a QH edge. Maxima
exhibited at $\nu=1$ follow two curves in $V$--$B$ parameter space
whose expression we give in terms of the lowest--Landau--level
dispersion. Their explicit form enables edge--dispersion spectroscopy.
At Laughlin--series filling factors, $\chi$LL behavior results in the
suppression of one of these maxima and characteristic power--law
behavior exhibited by the other one. The multiplicity of edge modes at
hierarchical filling factors corresponds directly to the multiplicity
of maxima in the differential tunneling conductance.

We thank M.~Grayson and M.~Huber for many useful discussions and
comments on the manuscript. This work was supported by DFG Grant
No.~ZU~116 and the DIP project of BMBF. U.Z.\ enjoyed the hospitality
of Sektion Physik at LMU M\"unchen when finishing this work.


\end{document}